\documentclass[twocolumn,twocolappendix]{aastex631}

\usepackage{amssymb,amsmath,amsfonts}

\newcommand{\rcd}{r_{\rm cd}} 
\newcommand{\rid}{r_{\rm id}}
\newcommand{\rt}{r_{\rm t}}
\newcommand{\rvir}{r_{\rm vir}}

\begin{document}


\shorttitle{Depletion boundary evolution}
\shortauthors{Gao et al.}

\title{Physical evolution of dark matter halo around the depletion boundary}

\author{Hongyu Gao}
\affil{Department of Astronomy, School of Physics and Astronomy, Shanghai Jiao Tong University, Shanghai, 200240, People’s Republic of China}
\affil{Key Laboratory for Particle Astrophysics and Cosmology (MOE), Shanghai 200240, China}
\affil{Shanghai Key Laboratory for Particle Physics and Cosmology, Shanghai Jiao Tong University, Shanghai, 200240, People’s Republic of China}

\author[0000-0002-8010-6715]{Jiaxin Han}
\affil{Department of Astronomy, School of Physics and Astronomy, Shanghai Jiao Tong University, Shanghai, 200240, People’s Republic of China}
\affil{Key Laboratory for Particle Astrophysics and Cosmology (MOE), Shanghai 200240, China}
\affil{Shanghai Key Laboratory for Particle Physics and Cosmology, Shanghai Jiao Tong University, Shanghai, 200240, People’s Republic of China}

\author{Matthew Fong}
\affil{Department of Astronomy, School of Physics and Astronomy, Shanghai Jiao Tong University, Shanghai, 200240, People’s Republic of China}
\affil{Key Laboratory for Particle Astrophysics and Cosmology (MOE), Shanghai 200240, China}
\affil{Shanghai Key Laboratory for Particle Physics and Cosmology, Shanghai Jiao Tong University, Shanghai, 200240, People’s Republic of China}

\author[0000-0002-4534-3125]{Y.P. Jing}
\affil{Department of Astronomy, School of Physics and Astronomy, Shanghai Jiao Tong University, Shanghai, 200240, People’s Republic of China}
\affil{Key Laboratory for Particle Astrophysics and Cosmology (MOE), Shanghai 200240, China}
\affil{Shanghai Key Laboratory for Particle Physics and Cosmology, Shanghai Jiao Tong University, Shanghai, 200240, People’s Republic of China}
\affil{Tsung-Dao Lee Institute, Shanghai Jiao Tong University, Shanghai, 200240, People’s Republic of China}

\author[0000-0001-7890-4964]{Zhaozhou Li}
\affil{Centre for Astrophysics and Planetary Science, Racah Institute of Physics, The Hebrew University, Jerusalem, 91904, Israel}

\correspondingauthor{Jiaxin Han}
\email{jiaxin.han@sjtu.edu.cn}

\begin{abstract}
    We investigate the build-up of the halo profile out to large scale in a cosmological simulation, focusing on the roles played by the recently proposed depletion radii. 
    We explicitly show that halo growth is accompanied by the depletion of the environment, with the inner depletion radius demarcating the two. This evolution process is also observed via the formation of a trough in the bias profile, with the two depletion radii identifying key scales in the evolution. 
    The ratio between the inner depletion radius and the virial radius is approximately a constant factor of 2 across redshifts and halo masses. The ratio between their enclosed densities is also close to a constant of 0.18. These simple scaling relations reflect the largely universal scaled mass profile on these scales, which only evolves weakly with redshift.
 The overall picture of the boundary evolution can be broadly divided into three stages according to the maturity of the depletion process, with cluster halos lagging behind low mass ones in the evolution. We also show that the traditional slow and fast accretion dichotomy of halo growth can be identified as accelerated and decelerated depletion phases respectively. 
\end{abstract}


\section{Introduction} \label{sec:intro}
With the developments of high-resolution numerical simulations, the dark matter distribution within halos have been extensively investigated. Both the Navarro-Frenk-White (NFW) profile \citep{1996ApJ...462..563N, 1997ApJ...490..493N}, whose slope changes from $-1$ to $-3$ from inside to outside, and the Einasto profile \citep[e.g.,][]{1965TrAlm...5...87E,2004MNRAS.349.1039N,2005ApJ...624L..85M,2006AJ....132.2685M,2008MNRAS.387..536G,2011MNRAS.415.3895L}, with a flatter inner slope, can reasonably describe the halo density profile at small or median scales. On the other hand, as an extended structure embedded in a diffuse large-scale environment, a concise definition of the boundary of a halo is more challenging, and has attracted more attention over recent years in both theories~\citep[e.g.,][]{2005ApJ...631...41T,2008MNRAS.388....2H,2008MNRAS.389..385C,2013MNRAS.430..725V,2014ApJ...789....1D,Zemp14,2021MNRAS.505.1195G} and observations~\citep[e.g.,][]{More2016,Baxter2017,2017ApJ...836..231U,2019MNRAS.485..408C,Deason2020, 2021ApJ...915L..18L,Fong22}.

Classical definitions of the halo boundary are established from the spherical collapse model~\citep{1972ApJ...176....1G}. In this model, a uniform spherical region in the Universe first expands till a maximum radius called the turnaround radius, after which it collapses over a freefall time and virializes within the so-called virial radius. The density of the virialized structure is predicted to be a fixed value times that of the background universe~\citep{1996MNRAS.282..263E,1998ApJ...495...80B}. Considering that the spherical collapse model is no longer valid when shell-crossing occurs, \cite{1984ApJ...281....1F} and \cite{1985ApJS...58...39B} derived the self-similar solution by scaling the trajectories of different shells with the characteristic time and length. The density profile predicted by the self-similarity solution shows a clear power-law form, with obvious caustic features that reflect the accumulation of material at the apocentres of different orbits. Since the infalling dark matter particles also have angular momentum and the halo shape deviates significantly from the ideal spherical shape \citep{2002ApJ...574..538J}, the caustic features on the density profile measured in the simulation are almost entirely smoothed out, except for the outermost one that is composed by the latest accretion material. The apogee of this outermost orbit is defined as the well-known splashback radius \citep{2014ApJ...789....1D,2014JCAP...11..019A,2015ApJ...810...36M,2016MNRAS.459.3711S,2017ApJ...841...34M,2017ApJS..231....5D,2017ApJ...843..140D}, which in practice corresponds to the steepest slope of the density profile. This feature can become more visible in the anisotropic density profile~\citep{Wang22}. Within the splashback picture, recent attempts have also been made to define halo boundaries based on the decomposition of material in phasespace into infalling and orbiting components~\citep[][]{Aung21, Diemer22a,Diemer22b,Garcia22}.  
	
	Recently, \cite{FH21} (hereafter FH21) proposed the concept of depletion radii to characterise the halo boundary from a new perspective. As a halo consumes material from its surroundings to grow inside, the separation between the growing region and the depleting environment provides a natural boundary definition, which is named the inner depletion radius, $\rid$. According to continuity, this radius is located exactly where the mass inflow rate peaks. Besides this, FH21 argues that the depletion of the environment outside $\rid$ can be observed as a trough in the bias profile, which is a rescaled version of the halo density profile. The location of the minimum bias is thus defined as the characteristic depletion radius, $\rcd$, reflecting the consequence of depletion. Based on these two interpretations, \citet{Fong22} and \citet{2021ApJ...915L..18L} have measured the characteristic and inner depletion radii in observations respectively, using weak lensing and satellite kinematics. 
	
	In contrast to the conventional virial radii that suffer from pseudo evolution~\citep{diemer2013pseudo,Zemp14}, the depletion radii characterise the physical evolution of halos on their outskirts by construction. 	Moreover, it is worth mentioning that the $\rid$ is an excellent match to the optimal halo exclusion radius proposed by \cite{2021MNRAS.505.1195G} in a halo model of the correlation function. This further supports that the depletion radius can be used as a physical halo boundary that self-consistently decomposes the cosmic density field into different halos, as recently implemented in \citet{Yifeng}.
	
	Different from the traditional steepest density slope location definition of the splashback radius, which typically corresponds to a 75\% containment radius of splashback orbits, this inner depletion radius can be interpreted as the radius enclosing a highly complete population of splashback orbits. As illustrated in \citet{FH21}, within this radius, the back-splashing particles counter-acts the infalling stream, leading to a decrease in the mean infall velocity, leaving the infall rate to peak at $\rid$. This is in agreement with \cite{Garcia23} who demonstrated that their halo exclusion radius is approximately located at the outer cut-off of the orbiting component of halo particles ~\citep{Aung21}. These interpretations all point towards a physical connection between $\rid$ and the halo exclusion radius.
	
	The dynamical interpretation of the depletion radii depends crucially on the evolutions of density and bias around a halo. Using the velocity profile at $z=0$, FH21 was able to deduce the evolution trend of the density profile qualitatively. In this work, we extend the study of FH21 by directly examining the evolution of the density, bias, and mass flow profiles in the simulation over a wide range of redshift ($z=0-5$). As we will show, the two radii stand out clearly in the evolution of not only the density profile, but also the bias profile, highly consistent with their depletion interpretation. We will also explore their roles in halo growth as well as their own evolution in detail.

	

	\section{Data}\label{data}
	\subsection{Simulation and the depletion catalog}\label{data_simulation}
	
	We use the same simulation as used in FH21, which is an $N$-body simulation from the CosmicGrowth \citep{2019SCPMA..6219511J} simulation suite run with a $\rm{P^3M}$ code~\citep{2002ApJ...574..538J}, adopting a $\Lambda$CDM cosmology with parameters $\Omega_{\rm{m}} = 0.268$ and $\Omega_{\Lambda} = 0.732$. A total of $3072^3$ dark matter particles are resolved with a box size of $600\,\mathrm{Mpc}\,h^{-1}$ per side, corresponding to a particle mass of $m_{\rm{p}} = 5.54 \times 10^8 M_{\odot}\,h^{-1}$.  The comoving softening length is $\eta = 0.01 \,\mathrm{Mpc}\,h^{-1}$. Halos are identified by the friends-of-friends algorithm (FoF) \citep{1985ApJ...292..371D} with a standard linking parameter $b=0.2$. The subhalos and their merger histories are identified by the \textsc{hbt+} code \citep{2012MNRAS.427.2437H,2018MNRAS.474..604H}\footnote{\url{https://github.com/Kambrian/HBTplus}}. Throughout the paper, we adopt the mass enclosed by a  sphere with a virialized overdensity  $\Delta_{\mathrm{vir}}$ according to the spherical collapse prediction \citep{1998ApJ...495...80B} as the default definition for halo mass $M_{\mathrm{vir}}$.  The corresponding halo radius is virial radius $r_{\mathrm{vir}}$.
	
	The FoF halo algorithm with a linking-length of $b=0.2$ is optimized for dissecting halos according to the virial radius. 
	Because the depletion radius is typically a factor of $\sim 2.5$ times the virial radius~\citep{FH21}, our FoF catalog contains halos that overlap in their depletion boundary, distorting the profiles around them on the depletion scale (see Appendix~\ref{app:overlap} for a case study). To avoid such complications, we remove any halo whose distance to a more massive neighbour is smaller than the sum of their estimated $\rcd$'s (i.e., $2.5 r_{\rm vir}$), to produce a depletion-radius based halo catalog. This \emph{depletion} catalog will be used for the majority of this work. To understand the influence of the depletion selection, however, we also show some results for the original FoF catalog in Appendix~\ref{sec:nonclean}.
	
	In addition to the above cleaning, we further limit our analysis to halos with more than 500 bound particles at $z=0$ to ensure sufficient resolution. For each of these $z=0$ halos, we track their evolution from $z=0$ back to $z\simeq 5$ along its main branch as resolved by \textsc{HBT+}, and extract their profiles out to $\sim 10{\mathrm{Mpc}}\,h^{-1}$ in physical radius. Our final sample contains only those halos whose main branches can be identified from $z=0$ to $z\simeq 5$. This tractability requirement removes $\sim 7\%$ and $0.04\%$ of haloes from the galactic and cluster sized halo samples which we study below, and has negligible influences on the profile evolutions. The most bound position of the central subhalo is chosen as the centre of a halo when measuring its profiles.

	\subsection{Identifying the depletion radii}\label{data_rd}                
	The depletion radii, along with the classical turnaround radius, can be identified in the bias and mass flow rate (MFR) profiles around halos. In principle, this can be done on a halo-by-halo basis. 
	To suppress noises, however, we choose to work with stacked profiles around a sample of halos of similar sizes, and further use second-order local polynomial interpolation to locate the extrema of the profiles.     
	
		\subsubsection{The inner depletion radius and feeding rate}
	The mass flow rate (MFR), $I_{\rm m}$, is defined as
	\begin{equation}
	    I_{\rm m}(r)\equiv 4\pi r^2 \rho(r) v_{\rm r}(r),\label{eq:MFR}
	\end{equation} 
	where $\rho(r)$ is the density profile and $v_{\rm{r}}(r)$ is the radial velocity profile including the Hubble flow. A negative radial velocity corresponds to infalling motion. 
	For convenience, we will also call the negative value of MFR as the Mass Inflow Rate (MIR), and the maximum MIR as the \textit{feeding rate} of a halo. 

According to the continuity equation,
	\begin{equation}
 	    \frac{\partial \rho(r)}{\partial t}+ \frac{1}{4\pi r^2}\frac{\partial I_{\rm m}(r,t)}{\partial r}=0,
 	\end{equation} 
 	the slope of MFR determines the growth rate of the local density. In regions with a positive slope, $\partial I_{\rm m}/\partial r>0$, matter falls with an increasing rate along the path, causing a net drop in the local density, i.e., $\partial \rho/\partial t<0$. On the other hand, in regions with a negative slope, the infall motion slows towards the center, causing matter to pile up. The minimum of $I_{\rm m}$ (or maximum of MIR) marks the exact transition between the two regions, and is defined as the inner depletion radius, $\rid$. 
	
	Besides, the turnaround radius $r_{\rm{t}}$ \citep{Pavlidou14, Tanoglidis15, Korkidis20} is also a meaningful characteristic radius in the spherical collapse model, as it marks the starting point of the infall zone around a halo, namely, where the radial velocity of matter is zero. This radius can also be identified in the MFR profile at its up-crossing point through $I=0$.
	\subsubsection{The characteristic depletion radius}
	
	The bias profile around an individual halo is defined as~\citep{Han19}
	\begin{equation}
	b(r) = \frac{\delta_{\rm{hm}}(r)}{\xi_{\rm{mm}}(r)}, \label{equ:bias_func}
	\end{equation}
	where $\xi_{\rm{mm}}(r)$ denotes the non-linear matter-matter correlation function, and $\delta_{\rm hm}(r)\equiv \rho(r)/\bar{\rho}-1$ is the overdensity profile around the halo.
	
	When averaged over a sample of halos, we recover the commonly used equation of the population bias
	\begin{equation}
	    \langle b(r) \rangle = \frac{\xi_{\rm{hm}}(r)}{\xi_{\rm{mm}}(r)},
	\end{equation} 
	where $\xi_{\rm hm}(r)=\langle \delta_{\rm hm}(r) \rangle$ is the halo-matter correlation function.
	
	
	The characteristic depletion radius, $\rcd$, is defined where the bias profile reaches its minimum on the intermediate scale. To precisely identify $\rcd$ in the bias profile, we perform a second-order local polynomial interpolation using the data point with minimum bias as well as its two adjacent points. Then $\rcd$ is determined as the position of the minimum value of this polynomial.

	\begin{figure*}
	\centering
	\includegraphics[scale=0.45]{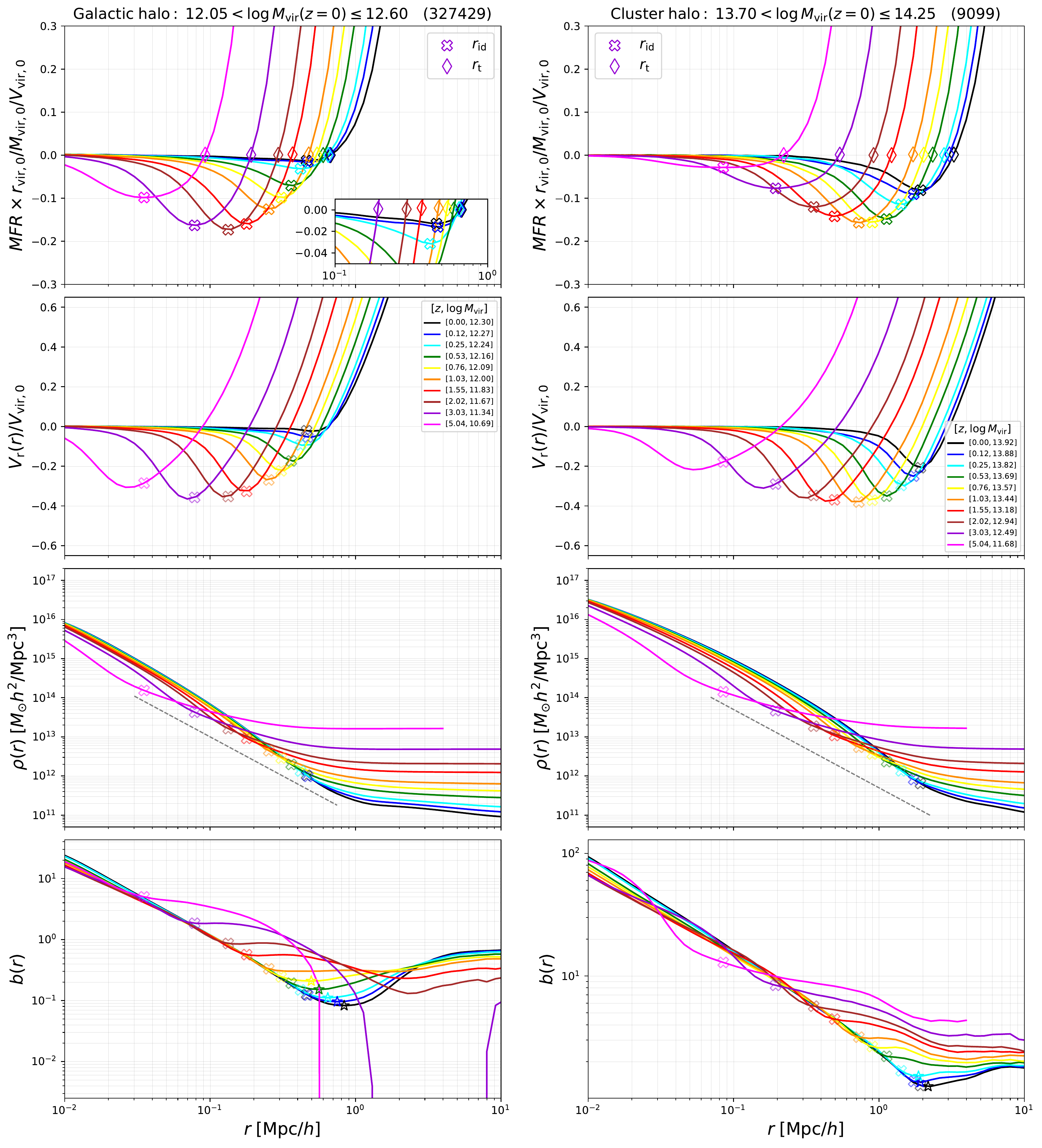}
	\caption{Evolution of MFR, radial velocity, density and bias profiles stacked in galactic-size halo bin (left column) and cluster-size halo bin (right column) are displayed in the four rows from top to bottom. The title of the top panel shows the number of the complete halo sample as well as the range of halo mass at $z=0$. Solid lines with different colors correspond to different redshifts, and the median halo mass at each redshift is also presented. The MFR profiles of the galactic-size halo bin are enlarged to show more clearly. The $\rid$, $\rcd$, and $\rt$ are denoted as different markers. The MFR and velocity profiles are all scaled by the virial quantities at $z=0$.  As a reference, we also show $\rho \left( r \right) \propto r^{-2}$ as a grey dashed line in the third panel.
	}
	\label{fig:1}
    \end{figure*}

	\section{Evolution of halo profiles in lights of the depletion radii}\label{evolution_halo}
	
	
	We select two samples of halos according to their virial mass at $z=0$, including a galactic-size halo sample with $M_{\rm vir}=10^{12.05}-10^{12.60} M_{\odot} \, h^{-1}$ and a cluster-size sample with $M_{\rm vir}=10^{13.70}-10^{14.25} M_{\odot} \, h^{-1}$. These halos are traced over time back to $z\simeq 5$ according to the \textsc{hbt+} merger tree. Figure~\ref{fig:1} shows the stacked profiles of the density, bias, MFR and radial velocity at a sequence of snapshots, along with various halo radii. 
	
	The MFR and velocity profiles in the top two panels of Figure~\ref{fig:1} all show a universal pattern. With an increasing radius from the halo centers, the radial velocity profile starts from a flat inner profile around zero, describing an approximately virialized inner halo, followed by a trough of negative velocity describing the infall of matter, and finally an outflowing outer region dominated by the Hubble expansion. As a halo grows, its depletion radii expand, and the corresponding troughs in MFR profiles move to larger scales. The evolution of the feeding rate, i.e., MIR at $\rid$, is not monotonic. It first increases and then decreases, peaking at $z\sim 2$ and $z\sim 1$ for the galactic and cluster size halos respectively. In particular, the mass accretion of galactic halos has almost completely stopped by $z=0$, with barely a trough in its MFR profile. 
 
  The shape of the radial velocity profile and its evolution are very similar to those of the MFR. Because the density profile is almost proportional to $r^{-2}$ on the $\rid$ scale \footnote{One should distinguish this scale from the scale radius $r_{\mathrm{s}}$ defined in NFW model, in which the slope of $\rho\left( r \right)$ reaches $-2$ at $r_{\mathrm{s}}$ and  continues to decrease to $-3$ as the radius increases. But at larger radius, the slope of $\rho\left( r \right)$ starts to increase and can reach $-2$ again around $\rid$.}, as illustrated by the grey dashed line in the third panel, the minimum of the MFR is close to the minimum of $v_{\rm{r}}(r)$ according to Equation~\ref{eq:MFR}. Therefore, although the MFR is an intrinsic description of the growth process of halos, the velocity profile can also be used to estimate $\rid$. Especially in observations, the radial velocity profile is more practical for measurements of the inner depletion radius \citep{2021ApJ...915L..18L}.
 
 	In the third panels we examine the density profile evolution in our simulation directly. Indeed the density evolves differently across $\rid$, with a growing inner profile and a decaying outer profile. The growth of the inner density also causes $\rid$ to grow over time. Given the finite time separation between the redshift bins, every two consecutive profiles cross each other in between their $\rid$'s. On the largest scale outside of $\rt$, the radial velocity is dominated by the Hubble flow, and the decrease in the density is primarily due to the expansion of the Universe. However, within the turnaround radius, the gravity of the halo becomes important, which can contribute significantly to the depletion of material outside $\rid$. 
  
As presented in the bottom panels, the evolution of the bias profile, where the average (over)density profile of the universe, $\xi_{\rm mm}$, has been scaled out, best reveals the consequence of the depletion process. In this representation, the large scale matter distribution becomes flat, allowing us to focus on the contribution from the halo itself to the density profile.  The characteristic depletion radius, $\rcd$, defined at the bias minimum, thus characterises the scale with the most prominent depletion signature. In the early evolution stage of halo,  matter around the halo has not been heavily depleted. The bias trough and its corresponding $\rcd$ are not formed at high redshifts. With a large amount of matter being fed into the inner part of halo within $\rid$, a clear bias trough can be seen in the bias profile outside $\rid$, which reflects the depletion of material due to the accretion of the halo.
 	
    The evolution of the bias profile clearly reveals the formation of the bias trough as a depletion process, in which the bias profile drops the most around $\rcd$. It is remarkable to see that each bias profile peels off from its progenitor right at $\rid$. This is highly consistent with the expectation that $\rid$ marks the inner edge of depletion, while $\rcd$ is located where the depletion is most significant.
 	
    An interesting phenomenon arising from this evolution process is that the bias profile out to $\rid$ is equivalent to the evolution path of $(\rid, b(\rid))$ at least approximately. In other words, the bias profiles at different redshifts are approximately universal within $\rid$, especially for the galactic size halos. This may be used to model the density profile evolution from the final bias profile and the $\xi_{\rm{mm}}$ evolution. We leave such an exploration to a future work.

	\section{Universal evolution of the outer halo}\label{evolution_rd}

	\begin{figure*}
	\centering
	\includegraphics[scale=0.6]{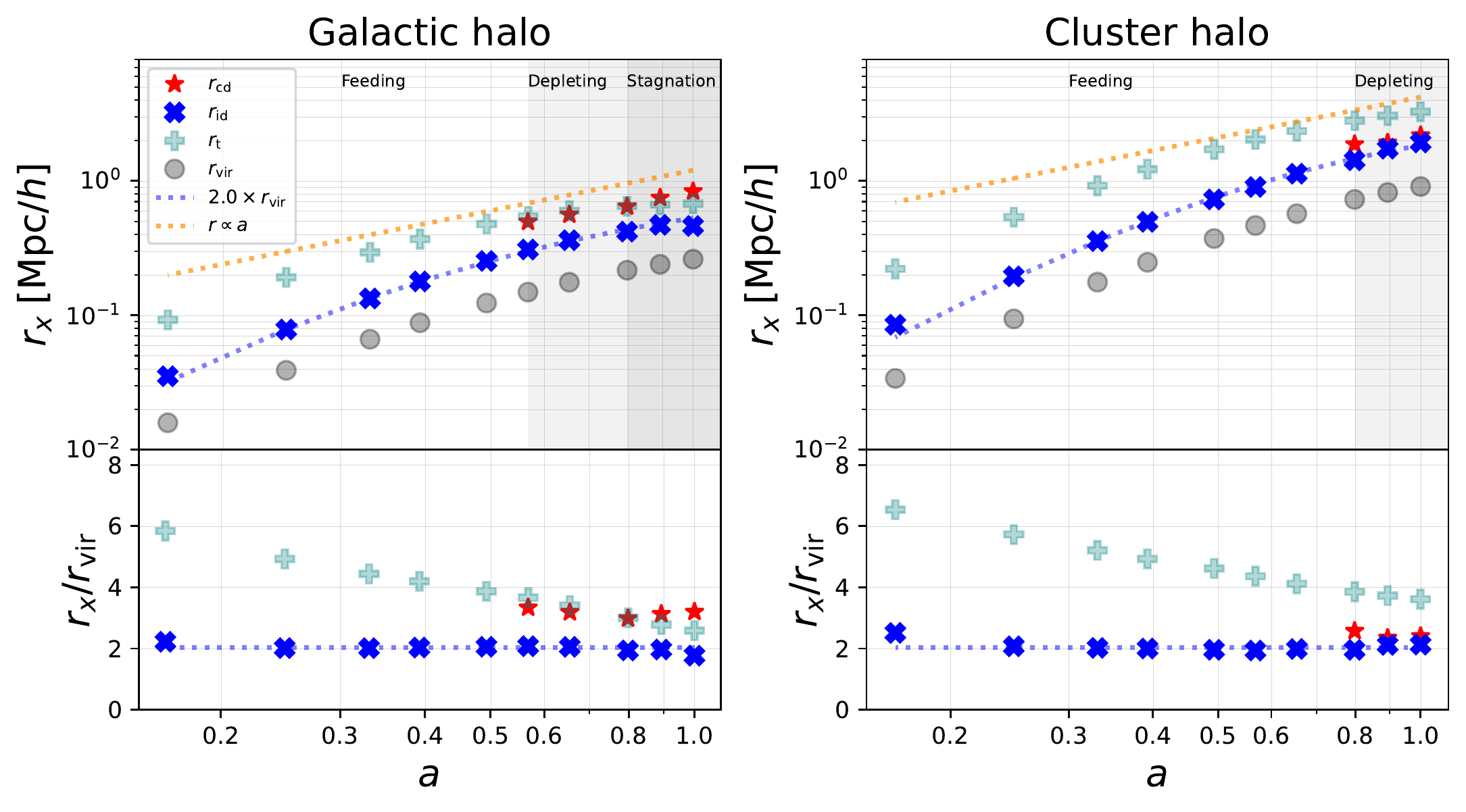}
	\caption{Evolution of depletion radii. The left and right panels correspond to the galactic-size and cluster-size bins in Figure \ref{fig:1}, respectively. $\rid$, $\rcd$, $\rt$ and $\rvir$ are marked as different symbols. 
         The three stages of the outer halo are divided by regions with different gray levels. The ratio of each radius to the virial radius is also shown in the bottom panel. We also display $2.0 \times \rvir$ as the blue dotted lines, which agree well with the $\rid$ at each redshift. To compare with the expansion rate of the universe, we plot $r\propto a$ as the orange dotted line. }  
	\label{fig:2}
	\end{figure*}

	\begin{figure}
	\centering
	\includegraphics[scale=0.6]{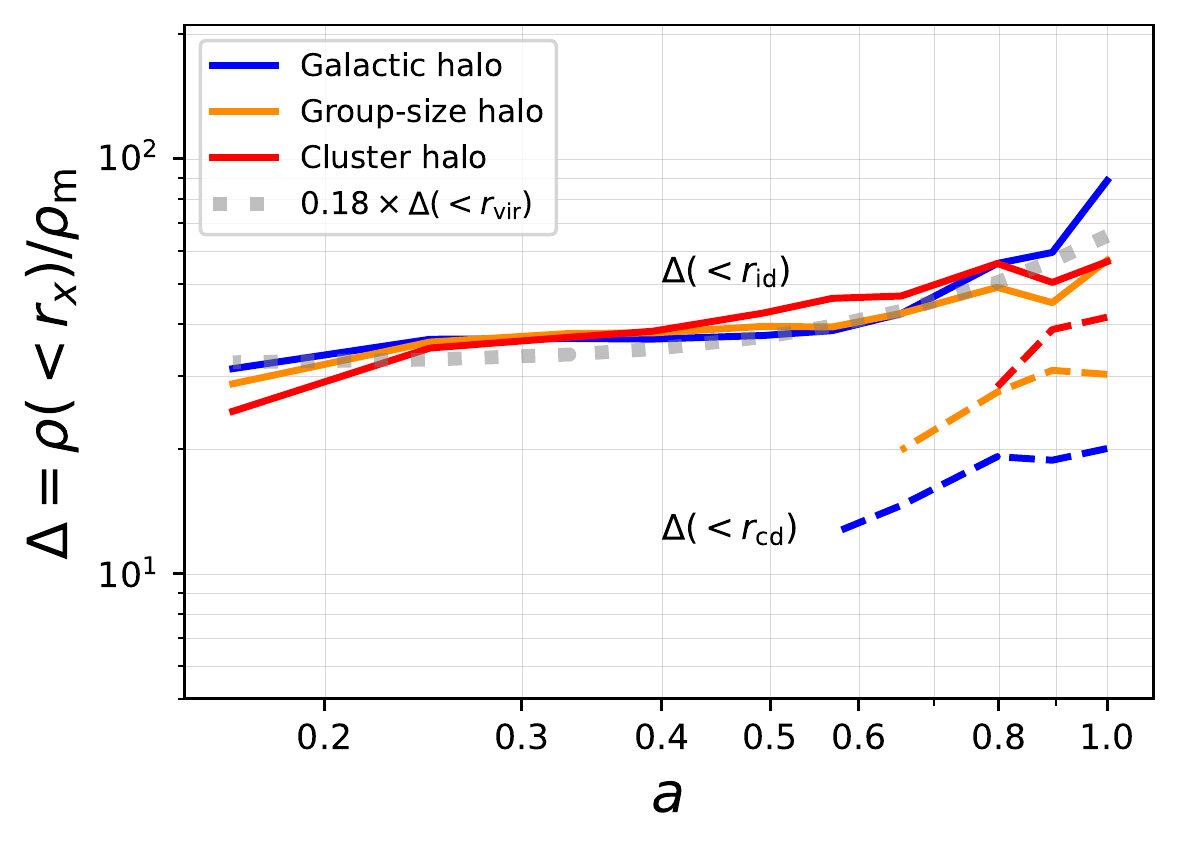}
	\caption{The evolution of density contrast of three halo samples with different mass. The solid and dashed lines  correspond to $\Delta \left(< \rid \right)$ and $\Delta \left(< \rcd \right)$, respectively. Different colours represent different mass. The dotted gray line represents $0.18 \times  \Delta \left(< \rvir \right)$.}  
	\label{fig:3}
	\end{figure}

	\subsection{Evolution of various halo radii}\label{trace_rd}

	Figure~\ref{fig:2} shows the evolution of various halo radii in the two halo samples. At high redshifts, it can be difficult to define $\rcd$ due to the absence of a bias trough. As a result, the evolution history of $\rcd$ is not complete in the figure. We only focus on those $\rcd$s that can be clearly identified. 
	
The overall shapes of the evolution histories of different radii are similar, with a slope that gradually flattens over time. 
	
At late times, $\rid$ grows faster than $\rt$. The depletion region enclosed by $\rid$ and $\rt$ will shrink as a halo evolves, eventually leading to the disappearance of the infall zone. Galactic halos have evolved further in this sequence than cluster halos, consistent with the late formation time of cluster halos in the virialized part. In section~\ref{sec:stages}, we will show that one can study the growth phases of halos both qualitatively and quantitatively according to these features.

We find that $\rid$ grows mostly in proportion to the virial radius across redshifts, following
\begin{equation}
	    \rid \simeq 2.0 \times \rvir. 
\end{equation} \footnote{Here the $\rvir$ is obtained by interpolating the stacked enclosed density profiles with $\Delta_{\rm{vir}}$. If we directly use the median $\rvir$ of the halo sample, this scaling relation is slightly changed to $\rid \simeq 2.1 \times \rvir$.} This scaling is consistent with the findings of FH21 at $z=0$. The good proportionality across redshift means that the growth rate of the outer halo is in pace with that of the virialized region. 

The growth of $\rcd$ appears more distinct. For galactic halos, $\rcd$ is well separated from $\rid$ in contrast to the close proximity of the two in cluster halos. This is different from the results of FH21 who found that the two radii are largely proportional to each other across mass. This difference can be explained by the depletion-radius based cleaning in our work, which primarily affects the low mass halos by excluding those in crowded environments, leading to a larger $\rcd$, while high mass halos are barely affected.
	
We have also plotted a reference $r\propto a$ curve in Figure~\ref{fig:2}, to compare the growth rate of the radii with the expansion rate of the universe. At early times, all the radii grow faster than the expansion of the universe. At late times, the growth rate of most radii have slowed down to roughly the same rate as the background expansion rate. For galactic halos, however, the turnaround radius has slowed down at an increased rate, leading to a much shrinked depletion region. The extra slowing down of the turnaround radius growth can be interpreted as caused by stronger tidal effects when the low mass halos become more clustered around massive ones, and the turnaround radius is the first to feel this effect as it is the outermost edge. We will carry out a detailed study on the effect of the large scale tidal field on halo growth in future work. The extra slowing down has also led halo growth into a distinct phase, as we will discuss in section~\ref{sec:stages}. Note that the proportionality of most radii to the scale factor at late time does not mean that halos are expanding freely on these scales. Instead, as we will show below, the enclosed densities within the various radii increase significantly relative to the background density at the late times.

	\subsection{Evolution of enclosed densities}
	In Figure \ref{fig:3} we show the evolution of density contrast enclosed by $\rcd$ and $\rid$, where $\Delta$ is defined as the ratio between the enclosed density and the mean matter density of the universe. In addition to the two mass bins studied above, we have further included an intermediate halo bin with $10^{12.75}<M_\mathrm{vir}<10^{13.30} M_{\odot} \, h^{-1}$. In general, the density contrasts increase with the expansion of the universe, reflecting that the halo stands out more clearly over time from the background on these scales. Moreover, the evolution of $\Delta(\rid)$ is approximately universal, which is mostly flat up to $a=0.5$ and grows faster afterwards. Intriguingly, such an evolution is well in proportion to the evolution of the virial density contrast, with 
	\begin{equation}
	    \Delta(\rid)\simeq 0.18\Delta(\rvir).
	\end{equation}
	This good proportionality indicates that the inner depletion radius may be modelled following the dynamics of spherical collapse. We will investigate such models in future work.

    Despite the overall proportionality to the virial density, the detailed evolution for the three mass subsamples still shows some interesting differences from each other especially at low redshifts.  
    
    To better understand this, we check whether there are other properties beyond halo mass that affect the depletion radii and densities. Indeed, as shown in Appendix~\ref{multiple}, the $\Delta \left(< \rid \right)$ at $z=0$ is more dependent on the halo formation time than on mass. The early formed halo exhibits a significantly high value of $\Delta \left(< \rid \right)$. Among our three halo samples, the galactic sized sample contains a higher fraction of early-forming halos. This explains its higher $\Delta \left(< \rid \right)$ at $z=0$ than the other two halo samples. Additionally, due to the very flat (shallow) troughs, the measurements of these $\rid$s and $\Delta \left(< \rid \right)$ at $z=0$ may be affected by some potential systematic errors such as the bin setting and fitting method.
	
	The evolution of the $\Delta(\rcd)$ shows an obvious mass dependence. This mass dependence is not observed in the $z=0$ FoF sample of FH21. This again can be explained by the extra depletion-radius based cleaning applied to our sample, which selects isolated low mass halos that have a lower density environment, while massive halos are barely affected. Without cleaning, the enclosed density within $\rcd$ is also found to be approximately universal, following $\Delta(\rcd)\simeq 44.77\times a^{1.79}$ over the redshift range where $\rcd$ can be identified (see Appendix~\ref{sec:nonclean}). However, as we discussed in section~\ref{data_simulation} and Appendix~\ref{app:overlap}, cleaning is a necessary process to obtain a self-consistent depletion-bound catalog, so we will still focus on results from the clean catalog.
	
\subsection{The scaled outer mass profile}\label{trace_enclosed_mass}

 \begin{figure*}
		\centering
	\includegraphics[scale=0.45]{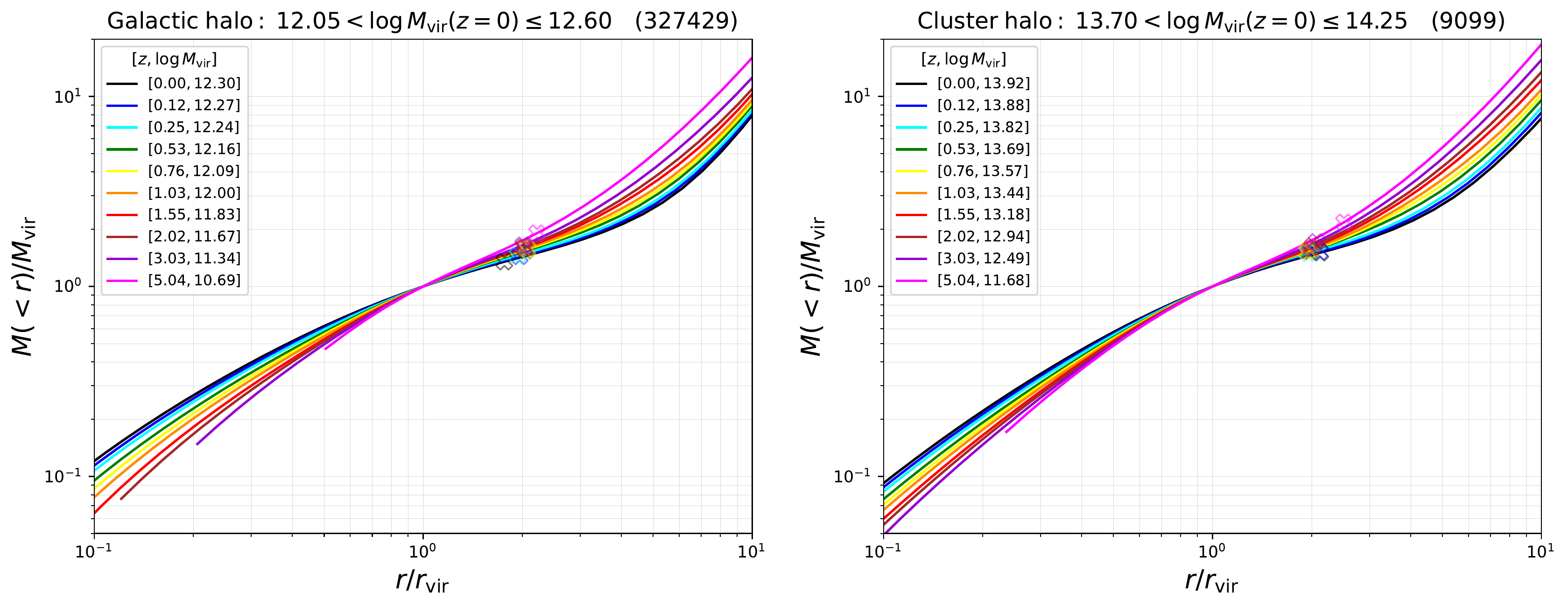}
	\caption{Evolution of  the scaled enclosed mass profiles. Each profile has been scaled by the virial radius $\rvir$ and  the virial mass $M_{\mathrm{vir}}$. The inner depletion radii $\rid$ are also marked with a cross on each profile. }
	\label{fig:scaled_enclosed_mass}
	\end{figure*}


  \begin{figure}
		\centering
	\includegraphics[scale=0.65]{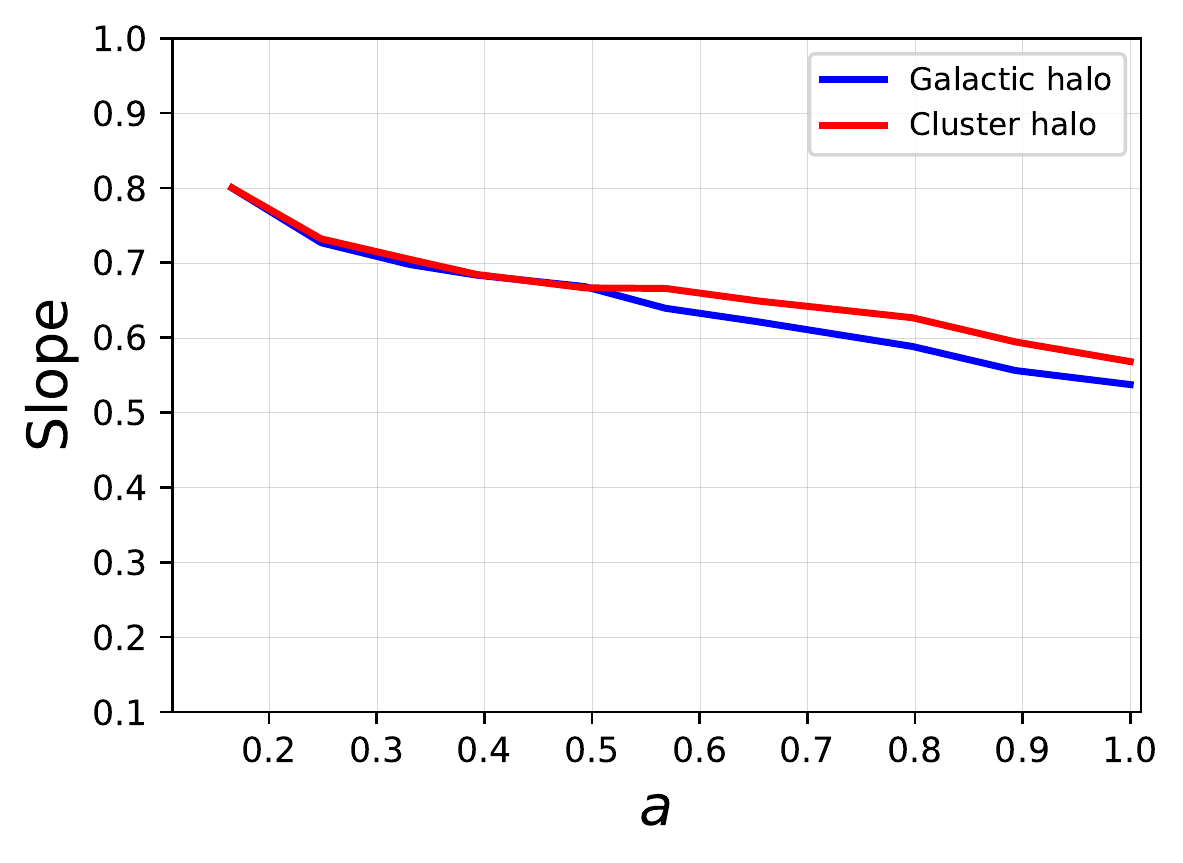}
	\caption{Evolution of  the power-law slopes measured from the scaled enclosed mass profiles (Figure ~\ref{fig:scaled_enclosed_mass}) between $\rvir$ and $\rid$. The blue and red colors correspond to the galactic halo and the cluster halo sample, respectively.}
	\label{fig:scaled_enclosed_mass_slope}
	\end{figure}


	
    The good proportionality of both the depletion radius and depletion density with the corresponding virial quantities suggests that the mass profile is universal in between the virial and the inner depletion radii, across mass and redshift. This is approximately the case shown in Figure~\ref{fig:scaled_enclosed_mass}, where the mass profiles between these two radii become largely unified when scaled by the virial quantities, for both galactic and cluster halos from $z=5$ to $z=1$. 
    
    In these coordinates, the depletion radii and the corresponding masses are tightly clustered around a single point. The two $z=5.04$ points and the $z=0$ point for the galactic halo appear as outliers, which can be attributed to the difficulty in accurately identifying their inner depletion radii from the corresponding kinematic profiles in Figure~\ref{fig:1}. 
   
    
    A power-law function of $M\propto r^{0.66}$ can describe the profile in between the two radii to an accuracy of $\lesssim 10\%$ for the redshift and mass ranges investigated. 
    As one expect the virial\footnote{Although the virial radius is expected to enclose a virialized structure that should no longer evolve, a realistic halo is not in complete equilibrium~\citep{oPDF2,oPDF3,oPDF4} and thus some density growth can still be observed within the virial radius in Figure~\ref{fig:1} (see \citealp{2008MNRAS.389..385C} for alternative definitions to the virial radius).} and inner depletion radii to bracket the physically growing part of a halo, such a unified power-law profile can be regarded as a manifestation of the similarity of halo growth~\citep{1984ApJ...281....1F,1985ApJS...58...39B}, although the power index $0.66$ is different from the asymptotic value of $\sim 1$ in the self-similar model.
    
    However, we emphasise that the unification of the profiles is only approximate. A weak but clear evolution of the scaled mass profile in this part is still present, suggesting the break-down of strict self-similarity. Quantitatively, the power-law index of the outer profile between $\rvir$ and $\rid$ decays slowly over time, from $0.8$ at $z=5$ to $\sim 0.6$ at $z=0$, as shown in Figure~\ref{fig:scaled_enclosed_mass_slope}. This deviation from exact universality also means that strict proportionality between $\rid$ and $\rvir$ can not hold simultaneously with strict proportionality between their enclosed densities, and some weak redshift evolution in the scalings is expected. More sophisticated studies of the profiles and of the inner depletion quantities will need to reflect this evolution. 
	
\section{Evolution phases of a halo}

\subsection{Three evolution stages of the outer halo}\label{sec:stages}
    By analysing the evolution of bias and MFR profile, we can phenomenologically divide the evolution of the outer halo into three stages.
	
    We take the evolution of a typical galactic-size halo as an example. The first stage is before the bias profile forms a clear trough (single minimum) on the depletion scale, 
    indicating that there is plenty of material to be fed to the halo. 
    At the same time, the MFR profile shows a prominent trough indicating that mass accretion and ongoing depletion is very active. We name this epoch as the ``feeding stage".\footnote{The bias goes to negative value on scales of a few $\rm{Mpc}$ for the two highest redshift bins. This is due to our selection of isolated halos on the depletion scale. The selection barely affects high mass halos, but tends to select low mass halos that are surrounded by under-dense regions at early time.} 
    
The feeding stage continues till $z\sim 0.76$, when a clear trough finally appears in the bias profile, representing a relative shortage of material in the halo neighbourhood. We name this stage as the ``depleting stage" during which the depletion is still active with a clear trough in the MFR profile, while the consequence of depletion is also visible as a bias trough. 

	With the depletion of the environment, the halo expands and the depletion radius becomes closer to the turnaround radius, leaving less and less material to be fed to the halo. This is accompanied with a decay in the feeding rate, which eventually approaches zero at $z\sim 0.12$, after which there is no obvious infall region around the halo. The growth of the halo almost halts, and we name this final stage as the ``stagnation stage".
	
	In the LCDM universe, more massive halos form later. Such a bottom-up structure formation paradigm is also reflected in the evolution phases of the outer halo. As shown in Figure~\ref{fig:1}, cluster halos transit to their depleting stages later ($z\sim 0.53$) than galactic halos ($z\sim 0.76$), and have not yet reached the stagnation stage as galactic halos have. 
	
 	The three evolution stages are distinguished with different levels of shading in Figure~\ref{fig:2}. Intriguingly, the depletion stages for both cluster and galactic halos start when the ratio between the turnaround radius and the inner depletion radius reaches slightly beneath $2$. For galactic halos, the stagnation stage starts when $\rt/\rid$ shrinks below $\sim 1.5$. This ratio measures the width of the active accretion zone, and depicts the richness of the environment that is feeding halo growth.
	
	\subsection{Connection to the inner growth phases}
\begin{figure}
	\includegraphics[width=0.5\textwidth]{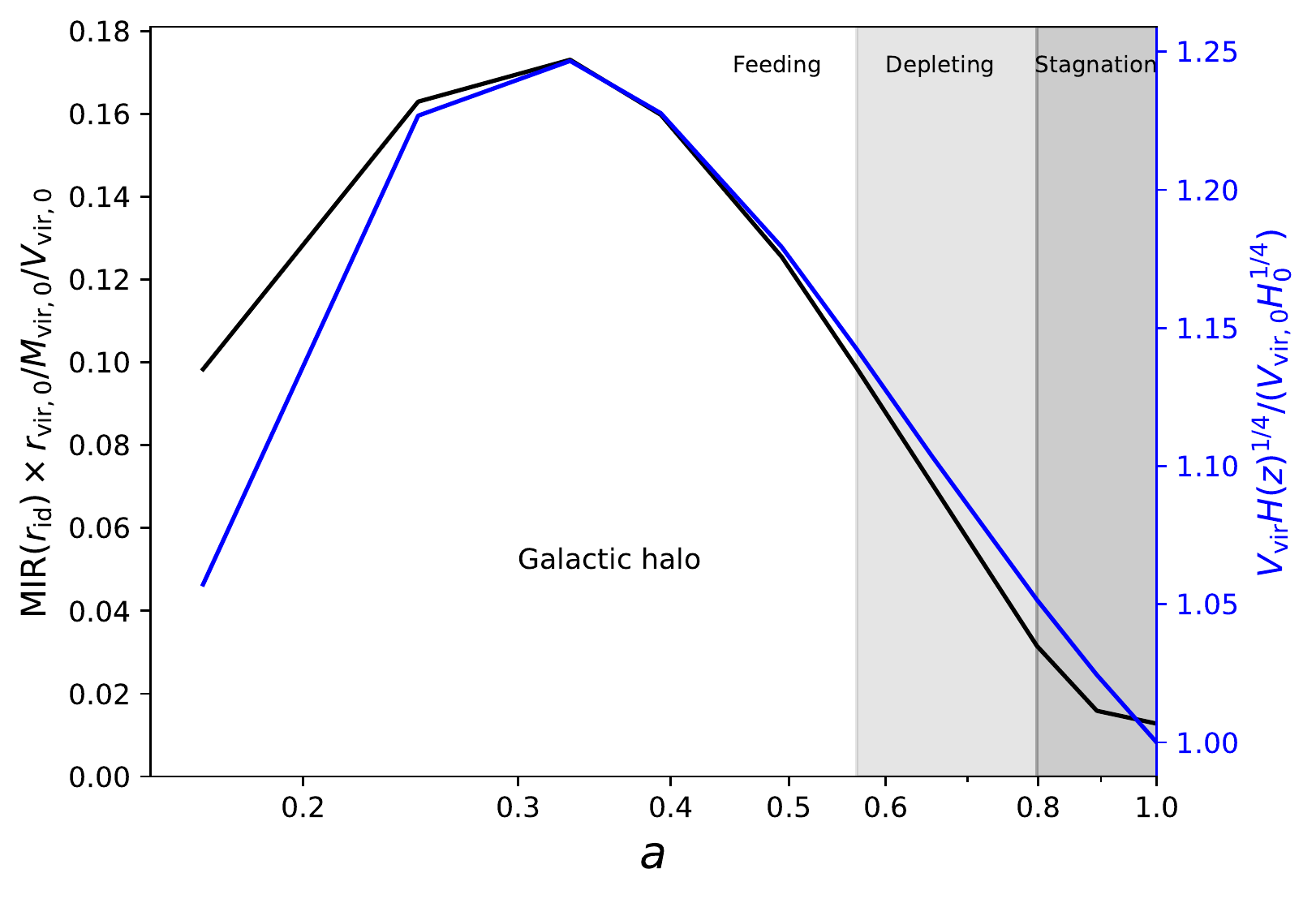}
	\caption{Evolution of the feeding rate, $\mathrm{MIR}(\rid)$, for galactic size halos, scaled by the virial quantities at $z=0$ (left axis and black line). The feeding rate peaks at around $a=0.35$ ($z\sim 2$). This coincides with the transition between fast and slow growth phases of the inner halo, which can be identified at the peak time of the $V_{\rm vir}H(z)^{1/4}$ evolution according to \citet{Zhao03}. The $V_{\rm vir}H(z)^{1/4}$ (right axis and blue line) is normalized by its $z=0$ value. The three evolution stages of the outer halo are also shown in different shadings.}\label{fig:peak_dep}
	\end{figure}

	\citet{Zhao03} proposed that the growth of the halo structure within the virial radius can be separated into a fast and a slow growth phase. In the fast phase, the halo mass grows faster than the Hubble flow and the inner structure of halo is mainly constructed during this phase. In the slow growth phase, the mass grows nearly in proportion to the Hubble rate and the mass enclosed by scale radius $r_{\rm{s}}$ has almost stopped growing. \citet{Zhao03} suggested that the separation of the two phases can be found at the time where $V_{\rm{vir}}H^{1/4}(z)$ peaks. We find that this transition time coincides with the time of the peak feeding rate, $\mathrm{MIR}(\rid)$. This is true for all the three halo mass samples that we examined, while in Figure~\ref{fig:peak_dep} we only show the galactic mass sample as an example. The transition between the fast and slow growth can also be observed in Figure~\ref{fig:1}, where the density grows relatively less significantly within $\rid$ after $z=2$ for the galactic halo.
	
	According to the evolution of the feeding rate, we can call the two phases alternatively as accelerated and decelerated depletion phases. The coincidence between the two peak times demonstrates that the evolutions of the inner and outer halos are in concert, and detailed studies of the accretion and depletion process in the outer halo could help us to better understand the evolution of the inner halo, and vice versa. In fact, even though the inner halo growth has been separated into two distinct phases, the growth rates of both the virial radius and the $r_{\rm s}$ evolve smoothly over time without a clear cut to divide the two. Thus \citet{Zhao03} have to propose a somewhat indirect proxy of $V_{\rm{vir}}H^{\gamma}(z)$ for separating the two phases, with $\gamma\sim 1/4$ being an empirical parameter. Our results suggest that the feeding rate defined in the outer halo can serve as a more objective and physical proxy for identifying halo growth phases even for the inner halo. 
	
        As the feeding rate describes halo growth at a different but physical scale, it is intriguing to explore its potential for quantitatively modeling the growth history of a halo. Following a similar line of thought, \citet{2021ApJ...915L..18L} attempted to utilize the ratio $\rid/\rvir$ as an indicator of the growth history of galactic halos. We leave such studies to future works.

	The transitions of the three stages of the outer halo evolution are also shown in Figure~\ref{fig:peak_dep}. These outer halo transitions all happen in the slow growth or decelerated depletion phase. This is consistent with the overall picture that halos grow from inside-out, with the inner part built-up on a smaller timescale than the outer part. 
	
	
	
	\section{Summary and Conclusions}\label{conclusion}
	
	The depletion radii have been proposed to describe the boundary of a halo according to the expected evolution of halo profiles in \citet{FH21}. By tracking halo evolution in an LCDM simulation, in this work we verify that the evolution of the density and bias profiles are indeed highly consistent with these expectations, with the growth of a halo accompanied by the depletion of its environment. The inner depletion radius, $\rid$, is shown to separate the growing part of the halo from a decaying environment, while also identifying the starting point of the depletion process in the bias profile. The characteristic depletion radius, $\rcd$, identifies the most depleted location in the bias, which is however only visible at late stages of halo growth.
	
	Both depletion radii expand with the growth of a halo. The evolution of $\rid$ closely follows the evolution of the virial radius for halos of a given mass, with $\rid\simeq 2\rvir$. Its enclosed density also evolves in proportion to the virial density, with $\Delta(\rid)\simeq 0.18\Delta(\rvir)$, irrespective of redshift or halo mass. These universal scaling relations are a consequence of the approximately universal mass profile in between these two radii. When scaled by the corresponding virial radii (or equivalently the depletion radii), the mass profiles between the two radii become largely identical across the redshifts and masses covered in this study, following a $M\propto r^{0.66}$ law, with residual evolutions observed at $\lesssim 10\%$ level. As these two radii are expected to bracket the physically growing part of a halo, the similarity of the profiles can be interpreted as a reflection of the approximate self-similarity of halo growth.
	
	For cluster halos, their $\rcd$ form relatively late and are close to the $\rid$s. For galactic halos, their $\rcd$ can be identified up to a higher redshift, which however are closer to their corresponding turnaround radius. Note the turnaround radii of these low mass halos are also intrinsically closer to their $\rid$ at low redshift, reflecting a lack of material to be fed to halo growth and the ceasing depletion process. 
	
	According to whether the $\rcd$ can be clearly identified and whether the depletion process is active, we can broadly divide halos into three evolution stages. These stages show different widths of their active accretion zones quantified by the ratio between the turnaround radius and $\rid$. Moreover, according to the evolution of the feeding rate, we can unambiguously divide halos into accelerated and decelerated depletion phases, which well correspond with the fast and slow growth phases of the inner halo known previously.
	
	These results illustrate the great potential of using the depletion radii as new probes for halo evolution. In a companion work~\citep{Yifeng}, we will also show that a more concise halo model can be built using these radii, which consists of a simple one-halo profile in the Einasto form and an intuitively scale-free halo-halo correlation. 
	Along with further theoretical understanding of them in analytical models, we expect many more applications of them can be found which can boost our understandings of halo evolution and structure formation in general.

\section*{Acknowledgments}
	We thank Yifeng Zhou and Xiaokai Chen for useful discussions. This work is supported by NSFC (11973032, 11890691, 11621303, 12133006), National Key Basic Research and Development Program of China (No.\ 2018YFA0404504), 111 project (No.\ B20019), and the science research grants from the China Manned Space Project (No.\ CMS-CSST-2021-A03). We thank the sponsorship from Yangyang Development Fund. We gratefully acknowledge the support of the Key Laboratory for Particle Physics, Astrophysics and Cosmology, Ministry of Education. ZZL thanks the support by ISF grants 861/20, 3061/21, and DFG/DIP grant STE1869/2-1 GE625/17-1. The computation of this work is partly done on the GRAVITY supercomputer at the Department of Astronomy, Shanghai Jiao Tong University.

\appendix
\section{The profile evolution when halos overlap on depletion scale}\label{app:overlap}

\begin{figure*}
	\centering
	\includegraphics[scale=0.5]{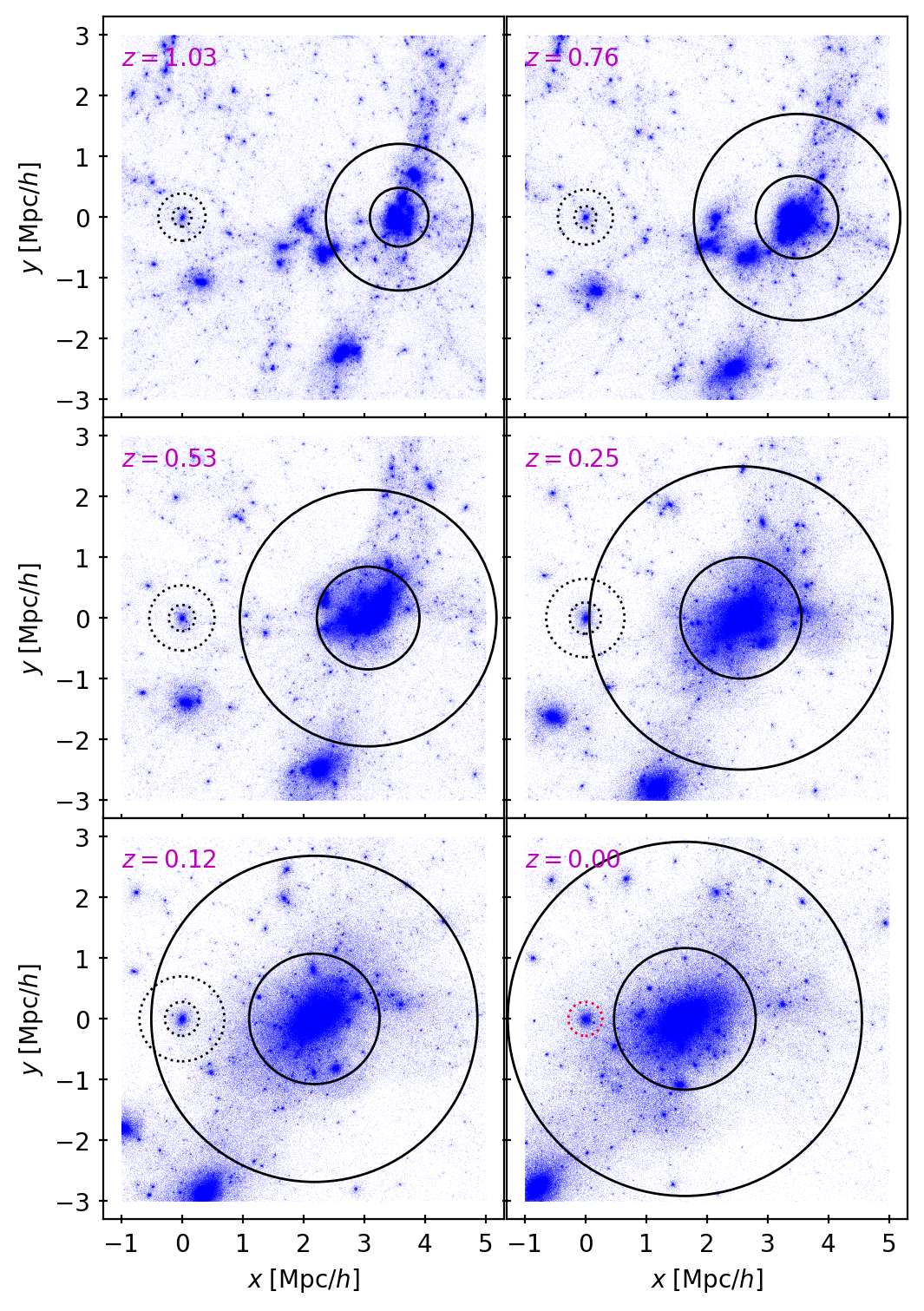}
        \hspace{0.5in}
	\includegraphics[scale=0.5]{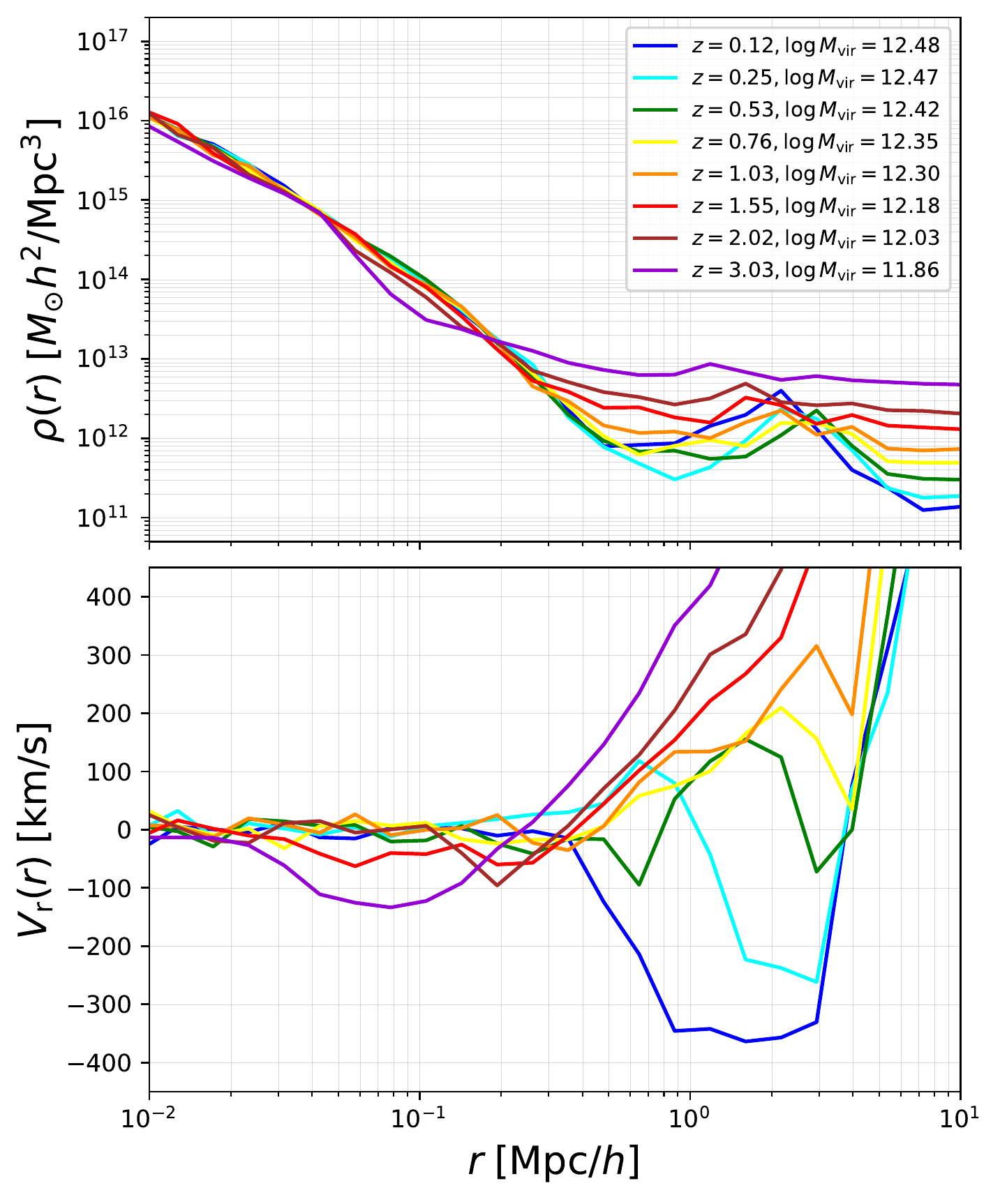}
	\caption{Left: Evolution and merger history of a pair of halos. The small halo is placed at the origin. The distributions of dark matter near the two halos are projected perpendicular to their alignment with a depth of 6 $\mathrm{Mpc}\,h^{-1}$. The black solid (dotted) inner circles represent the virial radii of the large (small) halo at different redshifts. The outer circle represents 2.5 times the virial radius, which is an approximate estimation of the characteristic depletion radius. The small halo eventually becomes a subhalo (highlighted with a red circle) of the large one at $z=0$. Right: Evolution of the density and radial velocity profiles around the small halo. 
	} \label{fig:merger}
\end{figure*}

In Figure \ref{fig:merger} we show the evolution history of a pair of halos that merge into a single halo at $z=0$. The density and velocity profiles are plotted from the center of the small halo. The evolution of the density profile is relatively simple, with the large halo appearing as a density peak outside the small one. The two halos are well separated initially, and a velocity profile typical of an isolated halo is observed at $z=3.03$, with a single trough reflecting the infall of matter towards the small halo. This trough becomes shallower and gradually shifts outwards as the halo grows. From $z\sim 0.5$, the velocity profile becomes very different from the expected form around an isolated halo, showing a peak within the trough before reaching the Hubble flow on large scale. This peak is due to the infall towards the large halo. At this time, the large halo is located at $r\approx3\mathrm{Mpc}\,h^{-1}$, forming the outer valley in the velocity profile. The material surrounding it fall with a larger velocity than the small halo, especially around the depletion radius of the large halo. These materials are seen as an outflow relative to the small halo, creating the velocity peak in between the two halos. As shown in the density profile, this process leads to the formation of a relatively low-density region at $r\approx1.5\mathrm{Mpc}\,h^{-1}$.
As the two halos further approach each other, $z=0.25$, the peak also moves to a smaller radius, and the outer trough further carves in reflecting the accretion by the large halo. After the small halo enters the depletion radius of the large one at $z=0.12$, the velocity profile is again dominated by a single trough, which however is due to the infall towards to large halo instead of the small one. 
Note at $z=0.12$ the two halos are still well separated outside their virial boundaries, while the infall regions of the two have merged. 

This reflects that the virial radius is not suitable for isolating halos when studying their evolutions out to the depletion scale. In fact, when using the depletion radius to define separate halos (outer circles in the left panel), the merger of the two halos starts soon after $z \sim 0.53$. After that the small halo can be no longer treated as an independent one, leading to complex structures in its velocity and MFR profiles.

This example illustrates the importance of defining halos self-consistently when studying the depletion radius. It could still be possible to study the depletion features for halos overlapping on the depletion scale. However, more careful treatments are needed to separate the overlapping objects and to account for the aspherical shape of the boundary, such as those done in \citet{2021ApJ...915L..18L}, which are not pursued in this work. 

\section{Evolution of the radius and density without cleaning}\label{sec:nonclean}

\begin{figure*}
	\centering
	\includegraphics[scale=0.6]{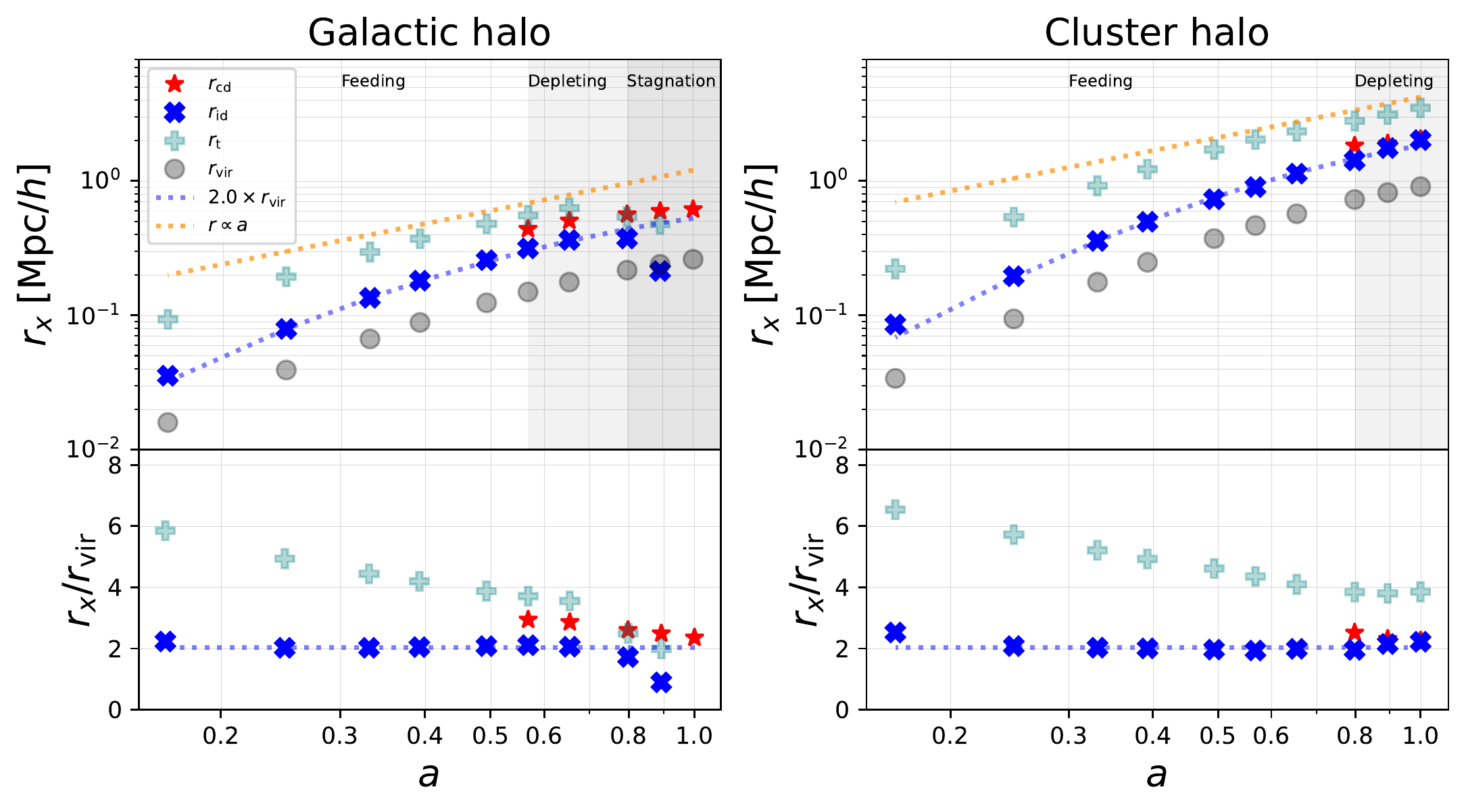}
	\caption{Similar to Figure \ref{fig:2}, but using the original FoF halo catalog before cleaning. Note the determination of $\rid$ can be problematic for galactic halos at low redshifts. }  
	\label{fig:evolution_radii_noclean}
	\end{figure*}

\begin{figure}
	\centering
	\includegraphics[scale=0.6]{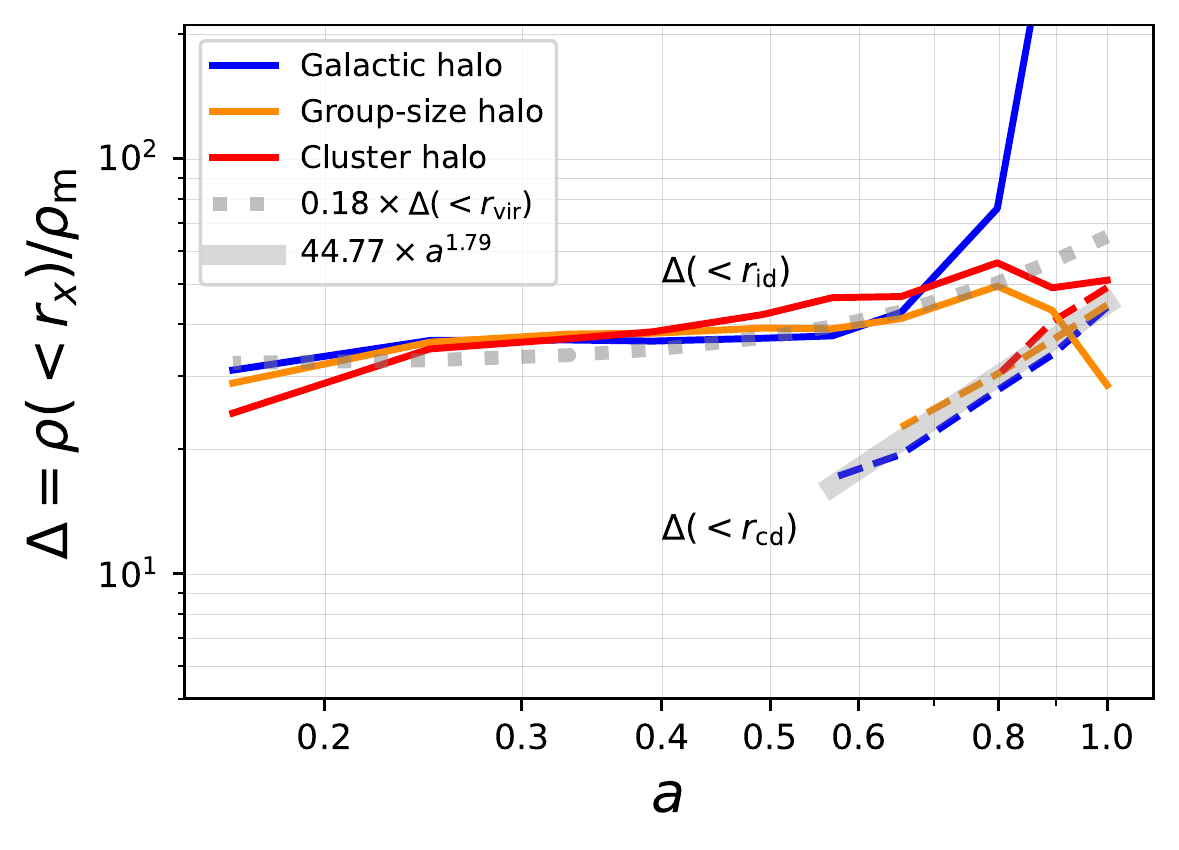}
	\caption{Similar to Figure \ref{fig:3}, but using the original FoF halo catalog before cleaning. Note the determination of $\rid$, hence $\Delta(<\rid)$, can be problematic for galactic halos at low redshifts.}  
	\label{fig:evo_delat_noclean}
	\end{figure}

In Figure~\ref{fig:evolution_radii_noclean} and Figure~\ref{fig:evo_delat_noclean} we show the evolution of the halo radii and densities extracted from the full FoF catalog without the depletion-radius based cleaning. For low mass halos, it becomes difficult to define their inner depletion radii due to the proximity to massive neighbours, as discussed in Appendix~\ref{app:overlap}. As a result, the $z=0$ measurements for $\rid$ are missing, and the low redshift results are generally ill-behaved. The $\rt$ results are affected similarly. Despite this, the $\Delta(<\rcd)$ results are more unified across mass. Note the high mass halos are barely affected by the cleaning.

\section{Dependence of the depletion radii on multiple halo properties at $z=0$}\label{multiple}


\begin{figure*}
\centering
\includegraphics[scale=0.35]{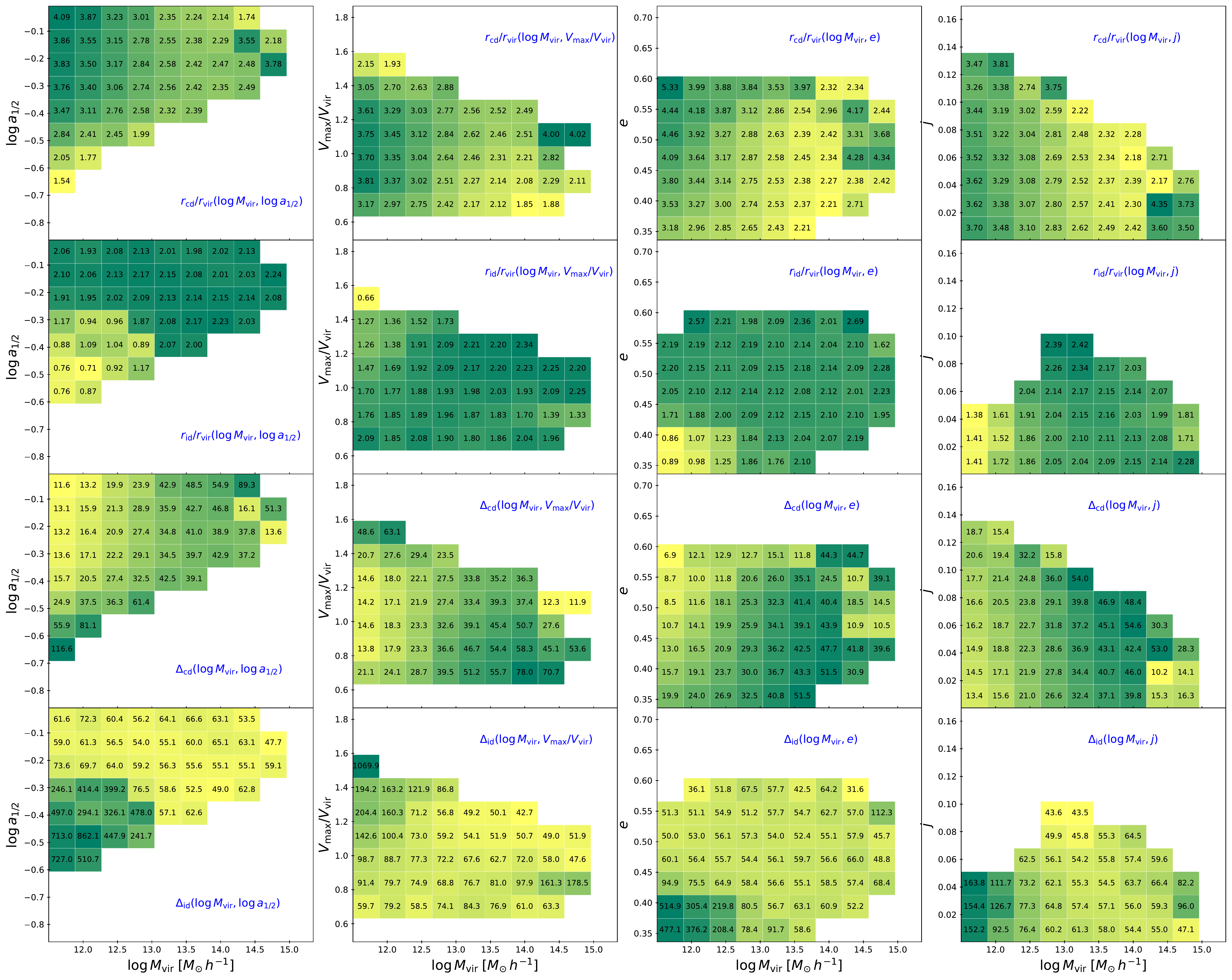}
\caption{Two-dimensional joint dependence of the depletion radius and density contrast on halo mass and other properties at $z=0$. Only the bins with more than 50 halos are displayed. The four rows from top to bottom represent the joint dependence of $\rcd$, $\rid$, $\Delta(<\rcd)$ and $\Delta(<\rid)$, respectively. The values of $\rcd$, $\rid$, $\Delta(<\rcd)$ and $\Delta(<\rid)$ are shown on each pixel. Here $\rcd$ and $\rid$ have been scaled by $\rvir$.}  
\label{fig:joint_dependence_scaled_by_rvir}
\end{figure*}

In Figure \ref{fig:joint_dependence_scaled_by_rvir}, we present the two-dimensional joint dependence of depletion radius $\rcd$ and $\rid$ as well as their enclosed density contrast  $\Delta(<\rcd)$ and $\Delta(<\rid)$ on halo mass $M_{\rm{vir}}$ and other halo properties at $z=0$. These physical properties include halo formation time $a_{1/2}$, concentration $V_{\rm{max}}/V_{\rm{vir}}$, shape $e$ and spin $j$. The detailed definition and calculation of these parameters can be found in \cite{Han19} and FH21. Only halos in the depletion catalog are used. 

These results extend the measurements in FH21, and show that the depletion properties are sensitive to many halo properties. Detailed studies on the origin, evolution and interplay of these dependences will be carried out in future works.




\bibliography{depletion_boundary}{}
\bibliographystyle{aasjournal}

\end{document}